\documentclass[fp,twocolumn,A4]{jpsj3}
\usepackage{txfonts}
\usepackage[dvipdfmx]{color}

\title{Structural evolution of a granular pack under manual tapping}

\author{Naoki Iikawa$^{1}$, Mahesh M. Bandi$^{2}$, and Hiroaki Katsuragi$^{1}$\thanks{E-mail address: katsurag@eps.nagoya-u.ac.jp}}
\inst{$^{1}$Department of Earth and Environmental Sciences, Nagoya University, Furocho, Chikusa, Nagoya, Aichi 464-8601, Japan \\
$^{2}$Collective Interactions Unit, OIST Graduate University, Onna, Okinawa 904-0495, Japan} 

\abst{We experimentally study a two-dimensional (2D) granular pack of photoelastic disks subject to vertical manual tapping. Using bright and dark field images, we employ gradient-based image analysis methods to analyse various structural quantities. These include the packing fraction ($\phi$), force per disk ($F_d$) and force chain segment length ($l$) as a function of the tapping number ($\tau$). The increase in packing fraction with tapping number is found to exponentially approach an asymptotic value. An exponential distribution is observed for both the cumulative numbers of force per disk $F_d$ : $N_{cum}(F_d) = A_F \exp (-\frac{F_d}{F_0})$, and force chain segment length $l$ : $N_{cum}(l) = A_l \exp (-\frac{l}{l_0})$. Whereas the coefficient ${A_F}$ varies with $\tau$ for force per disk, the force chain segment length shows no dependence on $\tau$. The $\tau$ dependence of $F_d$ and $\phi$ allows us to posit a linear relation between the total force of the granular pack ($F_{tot}^*(\tau)$) and $\phi(\tau)$.}

\begin{document}
\maketitle

\section{Introduction}
\label{sec:Introduction}
Granular packing is ubiquitous in everyday life. It is common knowledge that a denser granular pack can be achieved by tapping the pack. Clogged granular flow can be unjammed and structural foundations of buildings strengthened with tapping. Indeed, the first thing one does when in trouble with handling granular materials is to tap the container. Nevertheless, the physical mechanisms concerning the effect of tapping on granular packs are not yet completely understood. Recent investigations on granular compaction have used the dimensionless maximum acceleration $\Gamma=\alpha_{max}/g$ to characterize the strength of tapping and/or vibration applied to a granular pack, where $\alpha_{max}$ is the maximum acceleration and $g = 9.8$ m/s$^{2}$ is the gravitational acceleration~\cite{Katsuragi2015,Knight1995,Nowak1997,Josserand2000,Dijksman2009,Philippe2002,Philippe2003,Lumay2005,Lumay2006,Arsenovi2006,Ribiere2007}. Most previous studies have used steady vibration to cause granular compaction. The final state attained by steady vibration is solely determined by $\Gamma$ value~\cite{Knight1995,Nowak1997,Philippe2002,Arsenovi2006,Ribiere2007}. The most efficient compaction is induced at a value of $\Gamma \simeq 2$~\cite{Knight1995,Nowak1997,Philippe2003}. When $\Gamma$ is too small, the compaction takes long and grows logarithmically with time. When $\Gamma$ is too large, on the other hand, it is difficult to attain the highly compacted state as a large amount of kinetic energy is delivered to the granular pack in such a strong vibration~\cite{Philippe2003}. However, granular compaction also depends on vibration history~\cite{Josserand2000,Dijksman2009}. Although steady vibration has a well-defined maximum acceleration value, it represents one particular instance of granular compaction. In general, natural vibration or tapping applied to granular pack is somewhat irregular. Hence, granular compaction induced by irregular perturbations such as manual tapping must be examined to understand compaction processes that occur in nature. 

To diagnose the physical mechanism of granular compaction, an access to the inner stress structure created by granular pack is necessary. In general, the granular packs exhibit inhomogeneous stress distribution which can be characterized by a network of force chains. This force chain structure is peculiar to granular assemblies and causes their complex rheological behaviors. The force chain structure can be visualized in two-dimensional (2D) case. Using a 2D pack of photoelastic disks, the force chains can be observed via the retardation due to stress-induced birefringence of photoelastic material~\cite{Howell1999,Geng2001,Majmudar2005,Bandi2013,Puckett2013,Zheng2014,Behringer2014}. Using photoelastic disks, force applied to each disk can be measured~\cite{Howell1999,Geng2001}. More recently, the applied force has been decomposed into the normal and tangential components using a computational image matching method~\cite{Majmudar2005,Puckett2013}. Relations among the shearing, isotropic compression, and jamming have been intensively studied using photoelastic disks~\cite{Majmudar2005,Bandi2013,Puckett2013,Zheng2014,Behringer2014}. To the best of our knowledge, however, the tapping-induced granular compaction has not been studied with photoelastic disks. Therefore, we carry out the experiment with photoelastic disks to approach the physics of granular compaction via manual tapping. 

In this paper, we report the details of experimental methodology developed to study the granular compaction. A 2D granular pack consisting of photoelastic disks is constructed. Then, the manual taps are added to the granular pack. The evolution of packing fraction and force chain structure in the granular pack are characterized by image analysis of photoelastic disks. Using the analysed results, the relation between the packing fraction and force chain structure is discussed to reveal what happens in the compacted granular pack.

\section{Experiment}
\subsection{Setup}
\label{sec:Setup}

\begin{figure}
\begin{center}
\includegraphics[width=\hsize]{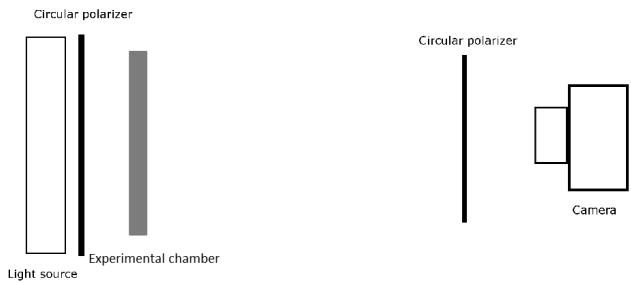}
\end{center}
\caption{Top view of the optical setup of the experiment. The experiment is carried out in a dark room to prevent stray light. The distance between the light source and the camera is about 1 m to make uniform angles of incident light into camera. The 2D experimental chamber is put vertically in front of the light source which is attached to a circular polarizer. A snapshot of the chamber is taken with a CCD camera (Nikon D70). The circular polarizer in front of the camera is set at $90^{\circ}$ (cross-polarisation mode) relative to the other. Two types of images are obtained with and without the circular polarizer in front of the camera.}
\label{fig:device_manual}
\end{figure}

\begin{figure}
\begin{center}
\includegraphics[width=\hsize]{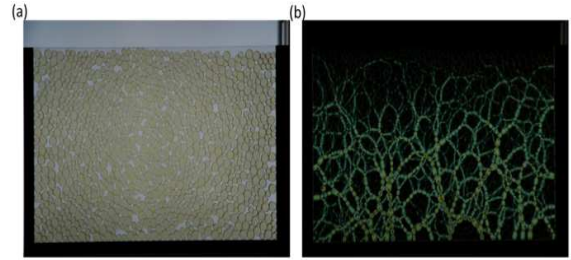}
\end{center}
\caption{(a) An example of bright field image by which the packing fraction and the position of photoelastic disks are obtained. (b) The corresponding dark field image by which the structure of the force chains is analysed}
\label{fig:Bright_and_dark_field_pictures}
\end{figure}

The experimental setup as shown in Fig.~\ref{fig:device_manual}, consists of a 2D experimental chamber constructed with acrylic plates along the front and back and held together by aluminium bars along the sides and bottom. The chamber dimensions are $0.3 \times 0.5 \times 0.011$ m in height, width and thickness, respectively. An accelerometer (EMIC Corp. Model: 710-C) is mounted on the top right corner of the chamber to measure the maximum acceleration ($\alpha_{max}$) experienced during the experiment. The chamber is filled with a bidisperse (to avoid crystallisation) set of 350 large (diameter is 0.015 m) and 700 small (diameter is 0.01 m) photoelastic disks of 0.01 m thickness (Vishay Micro Measurements, PSM-4). The chamber is vertically placed between a circular polarised LED light source and a CCD camera (Nikon D70), which acquires two types of images $2000 \times 3008$ pixels in size, corresponding to a spatial resolution of $1.76 \times 10^{-4}$ m/pixel (MPP). The camera is placed 1 m in front of the experimental chamber. First, a bright field image (Fig.~\ref{fig:Bright_and_dark_field_pictures}a) of the granular pack is acquired to measure the packing fraction and the disk centres and diameters for estimation of force per disk. A second dark field image (Fig.~\ref{fig:Bright_and_dark_field_pictures}b) is acquired by placing a second circular polariser between the experimental chamber and the camera in cross-polarisation mode. This image provides the photoelastic intensities of the granular force chains. Images are acquired under dark room conditions to minimise ambient noise from extraneous illumination.\\

\subsection{Experimental Protocol}
\label{sec:Experimental Protocol}
Prior to start of the experiment, an initial configuration of low packing fraction is generated. It is preferable that the initial packing fraction be small since this study focuses on granular compaction via manual tapping. However, when disks are introduced in a vertically standing chamber, initial compaction occurs from disk impacts. Therefore, the disks were introduced by spreading them in the chamber while it is horizontally laid down, and then the chamber was vertically fixed, thus assuring a small initial packing fraction. A pair of bright and dark field images is then acquired for this initial configuration.

The system is then perturbed by providing a manual tapping to the experimental chamber. In particular, each manual tapping is defined as adding two impulses to each bottom edge of the experimental chamber. Whereas this tapping protocol is not systematically controlled as in the case of an electromagnetic shaker, for instance, it was specifically chosen to mimic the situation of stochastic impulse forcing observed in many natural processes. In any event, the accelerometer attached to the experimental chamber measures the acceleration experienced during tappings, from which dimensionless acceleration is defined as $\Gamma \equiv \frac{\alpha_{max}}{g}$. The experiments reported here are in the regime of $\Gamma \simeq 3 - 4$. This tapping acceleration is large enough to achieve the efficient compaction. Following each manual tap, a pair of bright and dark field images is acquired for subsequent analysis to determine the evolution in the packing fraction ($\phi$), force per disk ($F_d$), and the force chain segment length ($l$) as a function of the tapping number $\tau$. Each experimental run consists of the initial configuration followed by nine manual taps ($\tau = 9$), thus providing ten pairs of bright and dark field images per run. Nine experimental runs under identical experimental conditions were conducted.

\subsection{Image Analysis}
\label{sec:Image Analysis}
Here we explain the image analysis methods employed to extract the packing fraction $\phi$, force per disk $F_d$, and force chain segment length $l$, from the bright and dark field images. The image analysis software, ImageJ \cite{abramoff2004}, was used to analyse the experimental image data.\\

\subsubsection{Determination of Packing Fraction}
\label{sec:Determination of Packing Fraction}
In this study, we define the packing fraction as $\phi = S_t/(S_m + S_v)$, where $S_t$ is the theoretical total area of the photoelastic disks, $S_m$ is the total area of photoelastic disks measured from the bright field images, and $S_v$ is the total void area measured from the bright field images. Whereas theoretically, $S_t = S_m$, in reality $S_m/S_t \simeq 1.1$ due to the optical distortion of the image between the centre and edges of the bright field image (see Fig.~\ref{fig:Bright_and_dark_field_pictures}a) and the thickness gap between disks and the chamber wall. When granular compaction occurs under manual tapping, whereas $S_m$ remains almost constant, $S_v$ decreases due to reduction in area of the voids between disks. Therefore, a measurable increase in packing fraction is observed with tapping number $\tau$. Often, the packing fraction is calculated as the ratio of the area occupied by the photoelastic disks to the total chamber area. This definition is reasonable when the granular pack is enclosed on all sides. However, since this experiment is conducted with the upper side of the experimental chamber left open, an accurate estimation of the total chamber area is not possible. The same situation also arises in estimation of packing fraction for granular heaps or sand piles \cite{Clement1992}.

For calculation of packing fraction $\phi$, the theoretical area of the disks $S_t$ was first calculated from the known number of large and small disks, whose diameters were already available, yielding $S_t = 1.17 \times 10^{-1}$ m$^2$ for experiments reported here. For calculation of $S_m$ and $S_v$, the bright field image was first binarized by using ImageJ which resulted in dark disks on a bright background. The pixel area of the dark regions multiplied by the spatial resolution (MPP) then provided the value of $S_m$, and inversion of images was used to obtain the value of $S_v$.\\

\subsubsection{Characterization of photoelastic intensity gradient}
\label{sec:Characterization of photoelastic intensity gradient}

\begin{figure}
\begin{center}
\includegraphics[width=\hsize]{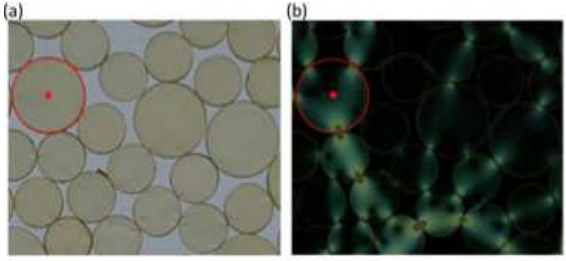}
\caption{Disk identification and measurement of force per disk from image analysis. The area and centre of each disk are obtained from the bright field image (a). Sample disk center and circumference are shown in red for a large disk. This information is then used in the corresponding dark field image (b) to obtain the photoelastic signal at disk contacts, and subsequent analysis is employed to measure the force per disk.}
\label{fig:Identification}
\end{center}
\end{figure}

Extant studies have used the photoelastic signal to measure contact forces in one of two ways. The first method estimates force per disk by using the photoelastic intensity gradient \cite{Howell1999, Zheng2014}. The second method estimates force per disk via computational image matching \cite{Majmudar2005, Puckett2013}. This study applies the former method for measurement of force per disk as the image resolution obtained is insufficient to measure forces by the latter computational matching scheme. The algorithm applied here for force measurement is similar to the one adapted by Howell et al. \cite{Howell1999}.

For the given intensity $I$ for each image pixel (8 bit, gray scale), Sobel filter was applied to obtain the squared horizontal $(\nabla I_h)^2$ and vertical $(\nabla I_v)^2$ gradients in intensity. Their sum $|\nabla I|^2 \equiv (\nabla I_h)^2 + (\nabla I_v)^2$ provides the squared gradient in intensity per pixel. The mean squared intensity gradient over all pixels within a disk was then defined as $\langle G^2 \rangle \equiv \langle |\nabla I|^2\rangle$. Computation of $\langle G^2 \rangle$ on each disk requires knowledge of each disk centre and its area; information available from the bright field image (Fig.~\ref{fig:Identification}) which proceeded in three steps: (1) binarizing a bright field image, (2) splitting disk areas of contiguous binarized intensity into individual disks, and (3) measuring each disk centre position and area. Step 1 is identical to the packing fraction measurement method. In step 2, a watershed algorithm was employed to discriminate between sharp gradients in intensity among disks, usually referred to as mountain (low intensity gradient) and river (high intensity gradient), and separates out individual disks. This is necessary to identify the disk perimeters along which the photoelastic intensities of contact forces exist.  Following this watershed procedure, each disk centre position and area were measured in step 3. Applying these results on the dark field image, the mean squared intensity gradient in photoelastic signal $\langle G^2 \rangle$ was then obtained for each disk.\\

\subsubsection{Force Calibration}
\label{sec:Force Calibration}

\begin{figure}
\begin{center}
\includegraphics[width=\hsize]{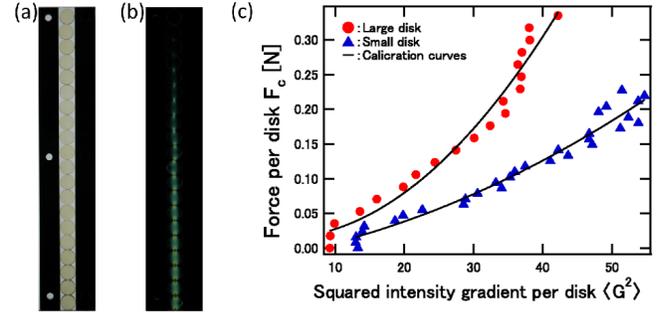}
\caption{Calibration method using a vertical 1D chain of photoelastic disks. Bright (a) and dark (b) field images of the vertical 1D chain. Image analysis of the 1D chain images using algorithms explained in section~\ref{sec:Characterization of photoelastic intensity gradient} provided values of $\langle G^2 \rangle$. Force per disk estimated from gravitational forcing $F_c$ was then used to relate $F_c$ versus $\langle G^2 \rangle$ for both large (solid red circles) and small (solid blue triangles) disks (c). Solid lines through the calibration data are quadratic fits, which were used in experimental measurement of force per disk $F_d$.}
\label{fig:Calibration_curve}
\end{center}
\end{figure}

The force per disk was calibrated using a vertical one-dimensional (1D) chain and the measurement method explained in section~\ref{sec:Characterization of photoelastic intensity gradient} to obtain force calibration curves which convert $\langle G^2 \rangle$ to force. A vertical 1D chain of photoelastic disks of height 0.3 m as shown in Fig.~\ref{fig:Calibration_curve}a \& b was constructed. The 1D chain consisted of either 20 large disks or 30 small disks. A pair of bright and dark field images was then obtained, and the image analysis methods (section~\ref{sec:Characterization of photoelastic intensity gradient}) were applied to calculate $\langle G^2 \rangle$ for each photoelastic disk. $F_c(n)$ in Newtons (where $n$ is the position of the disk from the top of the 1D chain), applied force per disk in the vertical 1D chain was estimated from the relation $F_c = n \times mg$, where $m$ is the mass per disk ($1.8 \times 10^{-3}$ kg for large disk and $0.8 \times 10^{-3}$ kg for small disk, respectively). In Fig.~\ref{fig:Calibration_curve}b, fringes on the boundary between disks and sidewalls cannot be observed in the dark field image. Therefore, the effect of sidewalls was neglected in calibration measurements. Figure~\ref{fig:Calibration_curve}c shows the calibration data obtained for both disk sizes.  The quadratic fits of the calibration data were then used as the final calibration curves for measurements of force per disk in the experimental data. Since the adopted procedure does not involve computational image matching of photoelastic fringes, only the total force applied to disk can be measured in this study, and cannot be decomposed into the normal and tangential components.\\

\subsubsection{Force chain segment length measurement}
\label{sec:Force chain segment length measurement}
\begin{figure}
\begin{center}
\includegraphics[width=\hsize]{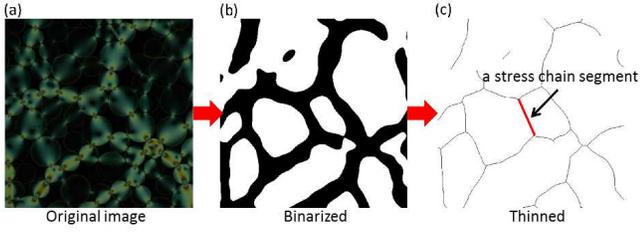}
\caption{Method of stress chain thinning and definition of stress chain segment length. A thinned stress chain (c) is obtained by binarizing original image ($(a) \rightarrow (b)$), and thinning it ($(b) \rightarrow (c)$). A stress chain segment length is defined by a liner distance between intersections or end points in thinned stress chain image.}
\label{fig:thinning}
\end{center}
\end{figure}

The segment length $l$ of force chains forms one of the structural variables measured in this experimental study. We employed a standard image analysis technique known as thinning method, an example of which is shown in Fig.~\ref{fig:thinning}. A dark field image (Fig.~\ref{fig:thinning}a) was binarized (Fig.~\ref{fig:thinning}b) and a skeletonize procedure (also known as the erosion method or the bleeding algorithm) in ImageJ was used to thin the segment down to a line of single pixel thickness. The force chain segment length was then defined as the linear distance between intersections or end points of the chain in the thinned force chain image (Fig.~\ref{fig:thinning}c).\\

\section{Results}
\subsection{Packing Fraction}
\label{sec:Packing Fraction}
\begin{figure}
\begin{center}
\includegraphics[width=\hsize]{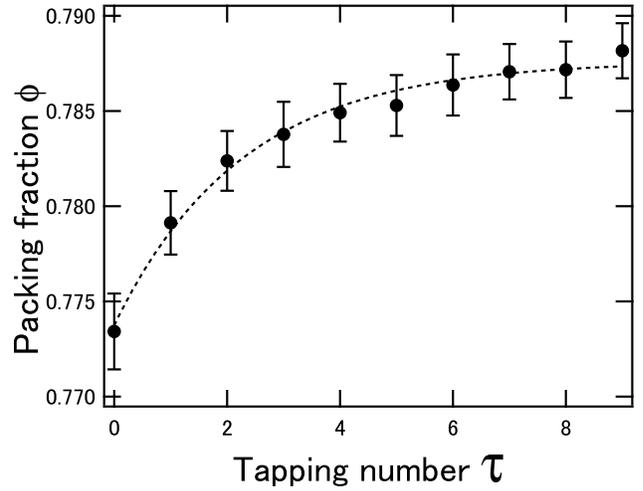}
\end{center}
\caption{The variation of packing fraction with manual tapping. Packing fraction increases with each manual tapping and approaches the steady state ($\phi_0 = 0.79$). Mean value of nine runs is reported, and the error bars represent the standard error of nine runs. The dotted curve is the fit, $\phi(\tau) = \phi_0 + A\exp(-\frac{\tau}{\tau_0})$, where $\phi_0 = 0.79$, $A = -1.39 \times 10^{-2}$ and $\tau_0=  2.27$ are values obtained for the fit parameters.}
\label{fig:packing_fraction}
\end{figure}

The calculated results for packing fraction at each tapping number $\tau$ are shown in Fig.~\ref{fig:packing_fraction}, where $\tau = 0$ represents the initial configuration. The experimental data for $\phi(\tau)$ are fit with the function $\phi(\tau) = \phi_0 + A\exp(-\frac{\tau}{\tau_0})$, where $\phi_0$, $A$ and $\tau_{0}$ are fit parameters. The fit parameter values for this study were found to be $\phi_0 = 0.79$, $A = -1.39 \times 10^{-2}$ and $\tau_{0} = 2.27$, respectively. As a result, Fig.~\ref{fig:packing_fraction} reveals that the packing fraction exponentially asymptotes towards a final steady state packing fraction value, in agreement with prior results reported by Bandi et al. \cite{Bandi2013}. Figure~\ref{fig:packing_fraction} reports the mean over nine experimental runs, with the error bars being their standard error.

\subsection{Force per disk}
\label{sec:Force per disk}
\begin{figure}
\begin{center}
\includegraphics[width=\hsize]{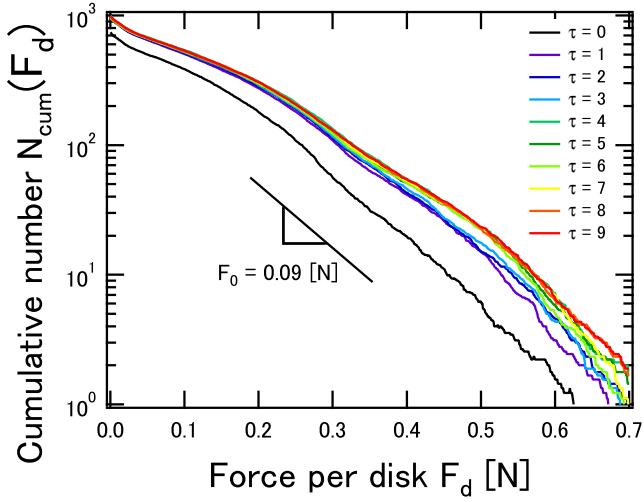}
\caption{Cumulative number distributions of force per disk at each tapping number $\tau$ in log-linear scale. The black solid line represents the initial configuration ($\tau = 0$) whereas colored lines represent the respective compacted states for various $\tau$ values. The data represent the mean value of nine experimental runs.}
\label{fig:Force applied to particle}
\end{center}
\end{figure}

\begin{figure}
\begin{center}
\includegraphics[width=\hsize]{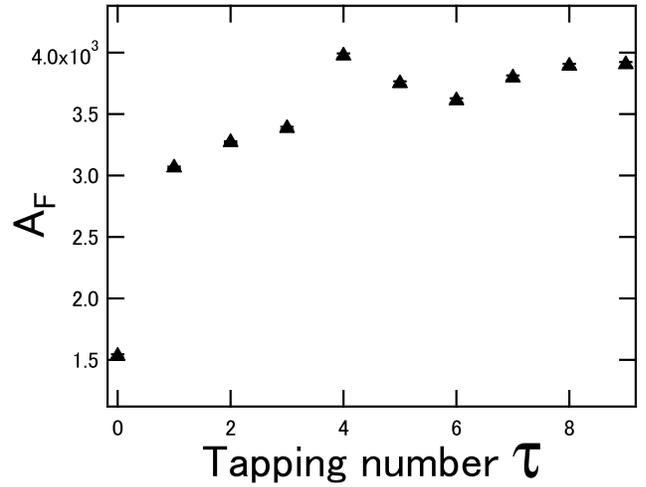}
\caption{The variation of fit parameter $A_F$ as a function of tapping number $\tau$. $A_F$ increases with each manual tapping. The $A_F$ values are calculated from the fitting by Eq.~(\ref{eq:forces applied to particle}) with fixed $F_0$ ( = 0.09 [N]). The error bars represent the uncertainty of the fitting.}
\label{fig:Af_variation}
\end{center}
\end{figure}

The force on each disk in the granular pack was measured using the method described in sections~\ref{sec:Characterization of photoelastic intensity gradient} \& \ref{sec:Force Calibration}. Figure~\ref{fig:Force applied to particle} shows the cumulative number distribution of force per disk at each tapping number $\tau$ in the granular pack. The range of force per disk $F_d$ in Fig.~\ref{fig:Force applied to particle} is wider than the calibration range (Fig.~\ref{fig:Calibration_curve}c). However, the calibration is performed under 1D diametral compression, i.e., coordination number is 2 in the calibration. In the granular pack, on the other hand, average coordination number is almost 4. Thus, the force applied to each disk can approximately be two times greater than the calibration case. Then, the force magnitude of each contact point in the granular pack is almost within the calibration range.The distribution can be approximated by the exponential form:
\begin{equation}
N_{cum}(F_d) = A_F\exp\left(-\frac{F_d}{F_{0}}\right),
\label{eq:forces applied to particle}
\end{equation}
where $N_{cum}(F_d)$ is the number of disks on which the applied force is equal to or greater than $F_d$. $A_F$ and $F_0$ are fit parameters with $F_0$ having dimensions of force. Figure~\ref{fig:Force applied to particle} is obtained from the mean value of nine identical experimental runs, and exhibits a roughly exponential distribution with almost constant slopes across all values of $\tau$, initial as well as the final compact state. The fit parameters were found to be $A_F = 1.53 \times 10^{3}$ and $F_0 = 9.06 \times 10^{-2}$ N at the initial configuration ($\tau = 0$). This result suggests that the functional form of the cumulative force distribution itself is invariant to the compaction under manual tapping as it yields the same slope for the exponential tail for all $\tau$ values. This result is qualitatively consistent with the previous study in which Liu et al. and Coppersmith et al. measured the cumulative distribution of force exerted by a three-dimensional (3D) granular pack on the container walls and showed that it follows an exponential distribution \cite{Liu1995,Coppersmith1996}.  Note that, however, the coefficient $A_F$ does vary with $\tau$ as shown in Fig.~\ref{fig:Af_variation}.\\

\subsection{Force chain segment length}
\label{sec:Force chain segment length}
Recent studies have analysed force chain segment lengths under pure shear and isotropic compression and found that they are exponentially distributed \cite{Peters2005, Sanfratello2011, Zhang2014.1}:
\begin{equation}
N_{cum}(l) = A_l\exp\left(-\frac{l}{l_{0}}\right),
\label{eq:a stress chain material}
\end{equation}
where $N_{cum}(l)$ is the number of segments of length equal to or greater than $l$. $A_l$ and $l_0$ are fit parameters and $l_0$ has dimensions of length. The unit of length used is the mean disk diameter $D = \frac{0.015 + 0.01}{2} = 0.0125$ m.

\begin{figure}
\begin{center}
\includegraphics[width=\hsize]{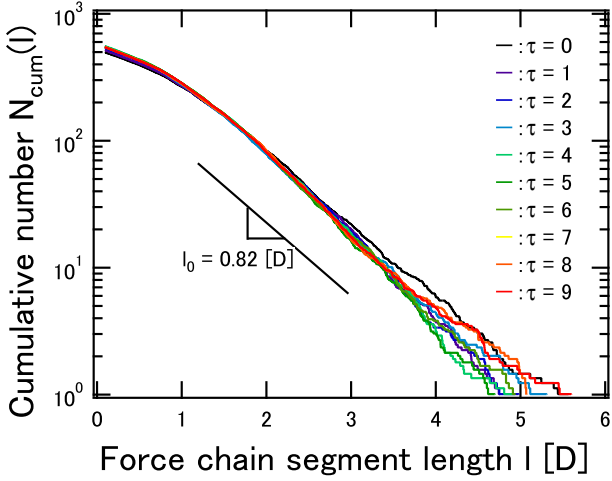}
\caption{Cumulative number distributions of force chain segment lengths in log-linear scale. The force chain segment lengths are quoted in units of the mean disk diameter, $D$. The black solid line represents the initial configuration ($\tau = 0$) whereas colored lines represent the respective compacted states for various $\tau$ values. The data represent the mean value of nine experimental runs.}
\label{fig:stress_chain_material_length}
\end{center}
\end{figure}

In agreement with previous works, the cumulative number distribution for the force chain segment length in this study is also found to be exponentially distributed (Fig.~\ref{fig:stress_chain_material_length}) with the functional form of Eq.~(\ref{eq:a stress chain material}). The fit parameters are $A_l = 9.45 \times 10^{2}$ and $l_0 = 0.82D \simeq 0.01$ m at the initial configuration ($\tau = 0$).  The characteristic length $l_0$ corresponding to the diameter of the small disk is derived from a mere reflection of the effect on analysis method. This fact indicates that a segment length is meaningless for less than the small disk size. This is natural because we consider the force chain structure consisting of disks. It is also clearly reflected in Fig.~\ref{fig:stress_chain_material_length} where the steady exponential slope is observed only for $l > 1D$. The slope and coefficient of the exponential distributions are almost constant across all $\tau$ values, rendering this distribution invariant to the manual tapping protocol.

\section{Discussion}
\label{sec:Discussion}
The experimental results discussed thus far show that the packing fraction varies with tapping number $\tau$ through the relation $\phi(\tau) = \phi_0 + A \exp(-\frac{\tau}{\tau_0})$, and saturating at an asymptotic value of $\phi_0 = 0.79$. Additionally, the cumulative distribution of force per disk at each $\tau$ was found to be exponentially distributed as: $N_{cum}(F_d) = A_F \exp(-\frac{F_d}{F_0})$. In particular, whereas the characteristic force $F_0$ remains invariant to $\tau$, the coefficient $A_F$ varies with $\tau$ (Fig.~\ref{fig:Af_variation}). On the other hand, the cumulative distribution of force chain segment length, which too is  exponentially distributed ($N_{cum}(l) = A_l\exp(-\frac{l}{l_{0}})$), exhibits no dependence on tapping number $\tau$ (Fig.~\ref{fig:stress_chain_material_length}). This suggests that the evolution of packing fraction $\phi(\tau)$ leads to the increment of internal force within the compacted granular pack, but does not lead to creation of new force chain segments. Following from these trends, we now explore a speculative relation between $\phi(\tau)$ and the total force $F_{tot}$ of the granular pack. Since $\phi$ is a globally averaged structural quantity, it ought to be compared with the total force.

\begin{figure}
\begin{center}
\includegraphics[width=\hsize]{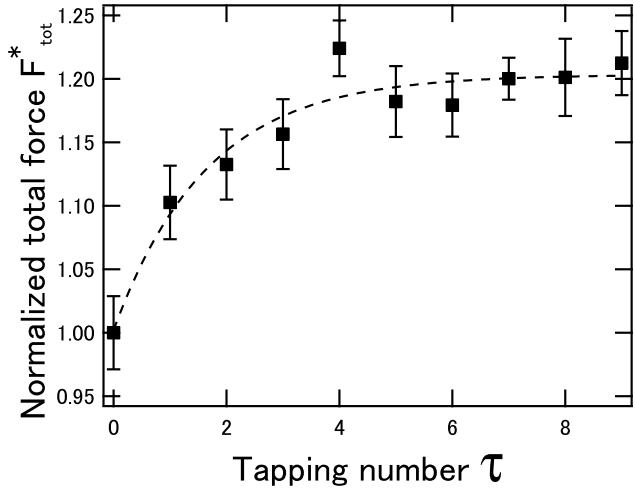}
\caption{Tapping number dependence of the normalized total force $F_{tot}^*$. Mean value of nine runs is reported, and the error bars represent the standard error of nine runs. The dashed curve is the fit, $F_{tot}^*(\tau) =  F_{t0}^* +  A_t^*\exp(-\frac{\tau}{\tau_{t0}})$, where $F_{t0}^* = 1.2$, $A_t^* = -0.2$ and $\tau_{t0}=  1.67$ are values obtained for the fit parameters. }
\label{fig:Ft_vs_tau}
\end{center}
\end{figure}

\begin{figure}
\begin{center}
\includegraphics[width=\hsize]{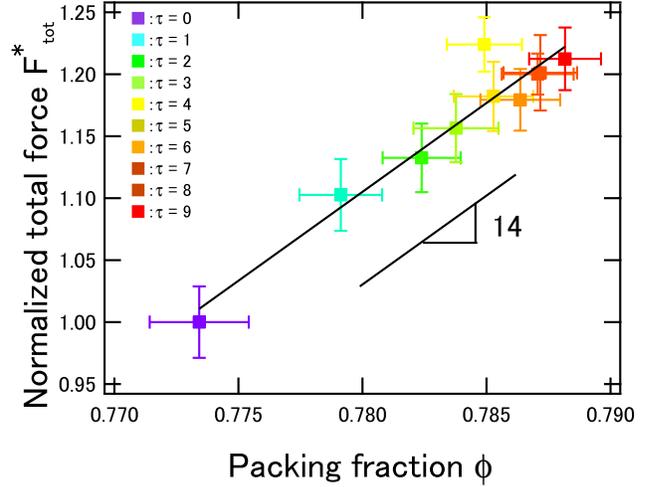}
\caption{Relation between $F_{tot}^*$ and packing fraction $\phi$. $F_{tot}^*$ is defined by $\displaystyle F_{tot}^* \equiv \frac{F_{tot}(\tau)}{F_{tot}(\tau =0)}$, where $F_{tot}$ is the sum of force per disk in the granular pack. Mean value of nine runs is reported, and the error bars represent the standard error of nine runs. The black solid line indicates the linear relation, $(F_{tot}^*(\tau)-F_{t0}^*)/(\phi(\tau)-\phi_{0}) = 14\pm2$.} 
\label{fig:Ft_vs_PF}
\end{center}
\end{figure}

The total force $F_{tot}$ is defined as $F_{tot} = \sum_{i=1}^k F_i$, where $F_i$ is the force per disk on the $i$th disk, and the summation is carried over all disks in the granular pack ($k$ represents the total number of disks), with the force threshold  placed at 0.1 N; forces below this threshold are not included in the summation. The total force is measured for the initial configuration and after each manual tap. Accordingly, we define a normalized total force as $F^*_{tot}(\tau) \equiv \frac{F_{tot}(\tau)}{F_{tot}(\tau=0)}$, where $F_{tot}(\tau = 0)$ represents the total force of the initial configuration. In Fig.~\ref{fig:Ft_vs_tau}, we show the normalized total force $F_{tot}^*$ as a function of $\tau$. We can confirm an asymptotic behavior of $F_{tot}^*(\tau)=F_{t0}^*+A_{t}^*\exp(-\frac{\tau}\tau_{t0})$, where $F_{t0}^*=1.2$, $A_{t}^*=-0.2$, and $\tau_{t0}=1.67$. This functional form is similar to that for $\phi(\tau)$. The comparison of $\phi(\tau)$ and $F_{tot}^*(\tau)$ reveals that $\tau_{t0} \simeq \tau_{0}$. Therefore, the ratio $(F_{tot}^*(\tau)-F_{t0}^*)/(\phi(\tau)-\phi_{0})$ should be approximated by $A_{t}^*/A = 14$. We independently confirm that $F_{tot}^*(\tau)$ vs $\phi(\tau)$ scales linearly as shown in Fig.~\ref{fig:Ft_vs_PF}. The slope of this scaling is $14\pm2$ in excellent agreement with the estimated result. This linear relation suggests that the process of compaction by tapping leads to the increment of granular internal force, in a linear fashion. This linear dependence may possibly result from the optimal amplitude perturbation ($\Gamma \simeq 3-4$) representing the linear response regime of the system. Stronger perturbations may not exhibit a similar dependence between $F^*_{tot}(\tau)$ and $\phi(\tau)$. This linear relation could be potentially useful for estimation of the increment of internal force within the compacted granular pack from packing fraction measurements for applications that involve compaction processes.

In this study, we developed the systematic methodology to analyse 2D granular pack comprising photoelastic disks. Using the developed method, granular compaction by manual tapping was analysed. Although the interesting structural evolution was revealed in this study, this is still the first step to approach granular compaction by tapping using photoelastic disks. The result should be compared with the case of controlled tapping using an electromagnetic shaker. Further studies concerning this comparison are in progress at present. The result will be published elsewhere.

\section{Conclusion}
\label{sec:Conclusion}
In this study, the structural evolution of 2D compacted granular pack has been experimentally studied using photoelastic disks. First, we developed the method to measure the packing fraction, contact forces, and force chain segment lengths using image analysis methods. Then, the dependence of these quantities on manual tapping were experimentally measured. From the experimental result, the exponentially asymptotic behavior of the packing fraction was observed. The distributions of applied force per disk and force chain segment length at each $\tau$ were characterized by exponential forms. Although the former depends on the tapping number $\tau$, the latter does not depend on it. The $\tau$-dependent total force is also shown to exhibit the asymptotic exponential behavior. The linear relation of these two functions ($\phi$ and $F_{tot}^*$) is confirmed from the measurements of $F_{tot}^*$ and $\phi$. 

\acknowledgments
The authors acknowledge S. Watanabe, H. Kumagai, S. Sirono, and T. Morota for fruitful discussions and suggestions. H. Katsuragi was supported by JSPS KAKENHI Grant Number 26610113. M. M. Bandi was supported by the Collective Interactions Unit at the Okinawa Institute of Science and Technology Graduate University.

\end{document}